\documentstyle[preprint,pre,aps,eqsecnum]{revtex}
\tightenlines
\begin{document}


\title{Phase-field model for Hele--Shaw flows 
with arbitrary viscosity contrast. I. Theoretical approach}

\author{
R. Folch, J. Casademunt, A. Hern\'andez--Machado}
\address{
Departament d'Estructura i Constituents de la Mat\`eria\\
Universitat de Barcelona,
Av. Diagonal, 647, E-08028-Barcelona, Spain
}
\author{L. Ram\'{\i}rez--Piscina}
\address{
Departament de F\'{\i}sica Aplicada\\
Universitat Polit\`ecnica de Catalunya,
Av. Dr. Mara\~n\'on, 50, E-08028-Barcelona, Spain
}

\maketitle

\begin{abstract}

We present a phase-field model for the dynamics of the interface 
between two inmiscible fluids
with arbitrary viscosity contrast in a rectangular
Hele--Shaw cell. With asymptotic matching techniques we check
the model to yield the right Hele--Shaw equations in the
sharp-interface limit and compute the corrections to these equations
to first order in the interface thickness. We also compute the effect of
such corrections on
the linear dispersion relation of the planar interface.
We discuss in detail the conditions on the interface thickness 
to control the accuracy and convergence of the phase-field model to
the limiting Hele--Shaw dynamics.
In particular, the convergence appears to be slower for 
high viscosity contrasts.

\vskip 5mm
\centerline{Copyright 1999 The American Physical Society}
\end{abstract}
\newpage

\section{Introduction}

The dynamics of morphologically unstable interfaces is a major
problem in nonequilibrium physics from both fundamental and applied
points of view. Relevant examples of those are dendritic growth,
directional solidification, flows in porous media, flame
propagation, electrodeposition or bacterial growth \cite{general}. 
The so-called Saffman-Taylor problem has played a central role in
this context because of its relative simplicity both experimentally
and in its theoretical formulation \cite{general,st}.
It deals with the motion of the
interface between two inmiscible fluids within a Hele-Shaw cell. 
Due to the highly nonlinear and nonlocal nature of the interfacial
dynamics of such systems, analytical understanding is scarce and restricted
to high viscosity contrast\cite{Siegel}, so in general 
one relies mostly on numerical work
\cite{aref1,aref2,aurora,jaume,vinals,shelley}.

     From a mathematical point of view, such systems are referred
to as moving boundary problems. In practice this implies that one
has to keep track of the interface where boundary conditions are
applied, and solve a (linear) problem in the bulk which determines
in turn the motion of the boundary. This kind of problem has
traditionally been addressed in terms of boundary integral methods
which reduce the dynamics of the interface to integrodifferential
equations. The numerical integration of these equations is
quite involved though, particularly for long times, due to stiffness and
numerical instability of the equations.
In the case of Hele-Shaw flows, boundary integral methods 
have succesfully been applied \cite{aurora,jaume,vinals,shelley},
although quite sophisticated
algorithms have usually been necessary \cite{shelley}. 

Recently, the so-called phase-field equations have been proposed in
the context of solidification problems as a different approach to
the interface dynamics 
\cite{langer,kobayashi,mcfadden1,karma,mcfadden2,grant,heine,mozos,ricard,caginalp}.
In the spirit of the well known
time-dependent Ginzburg-Landau models \cite{hh},
the method avoids the
tracking of the interface by introducing an auxiliary field
(analogous to an order parameter)
which locates the interface and whose dynamics is coupled to the
other physical fields through an appropriate set of partial
differential equations. In this way, there is no boundary condition
to explicitely apply at the interface and the whole system is
treated as bulk.

This method introduces a mesoscale $\epsilon$ which is
not present in the original macroscopic equations and gives a
finite thickness to the interface. The equations are then chosen in
such a way that the original
bulk equations and boundary conditions are recovered
in the $\epsilon \rightarrow 0$ limit. Therefore the
phase-field equations for a given model are not intended to
describe the true mesoscale physics of the system, and are then
not unique. In fact there is considerable freedom in choosing a
particular form of them, with criteria of either numerical
efficiency and convergence \cite{karma} or
other physical criteria such as
thermodynamic consistency \cite{mcfadden2}. In any case,
the nature of the
phase-field approach is completely different from the
sharp-interface models and therefore the actual numerical
advantages and limitations of both are also quite distinct. This
makes the two approaches complementary and competitive in different
physical situations. 
A remarkable advantage of the phase-field approach is that it is
much simpler to implement satisfactorily from a numerical point of
view. On the other hand, the phase-field approach is usually more
amenable to generalization, in the sense that it allows to
introduce variations and new elements without any major
modification of the numerical scheme, for instance in the treatment
of fluctuations, liquid crystals \cite{ricard}
and other complex fluids\cite{shelley}. Finally, the
phase-field approach can handle very naturally situations where the
sharp interface model is not appropriate, such as for instance 
topology changes like interface pinching leading to the breakup of 
bubbles.

In this paper we introduce a phase-field model for Hele-Shaw
flows with arbitrary viscosity contrast (or Atwood ratio)
$c=\frac{{\mu_1}-{\mu_2}}{{\mu_1}+{\mu_2}}$. 
Although in the high contrast limit $c=1$ the Hele-Shaw dynamics is
quite analogous to the one-sided solidification problem (in the
appropiate approximations \cite{vinals}), the arbitrary viscosity contrast case
has been shown to exhibit quite different dynamics than
solidification problems, and has in fact opened some interesting
questions, particularly concerning the sensitivity of finger
competition to viscosity contrast \cite{aref1,aref2,aurora,jaume,maher}
and the long time asymptotics of
the low viscosity contrast limit \cite{jaume}. 

The model presented here is inspired in the vortex-sheet
formulation of the problem \cite{aref1},
in which the relevant dynamic variable
in the bulk is the stream function. Similar ideas have previously been
applied to describe physically diffuse interfaces in the
context of steady state selection in thermal plumes \cite{benamar}. 
Usually phase-field models are naturally suited to symmetric
situations (two-sided models). The present case of Hele-Shaw flow
is no exception and becomes most efficient for $c=0$. 
Finite $c$ can also be handled but the model becomes
computationally inefficient in the limit $c\rightarrow 1$, since
this limit must be taken formally after the $\epsilon \rightarrow
0$ limit. A phase-field model for this one-sided case must differ
essentially from the one presented here, such as in the spirit of
Ref. \cite{caginalp}. 

The layout of the rest of the
paper is as follows: in Sec. \ref{secmac} we recall the 
Hele--Shaw macroscopic equations in terms of the stream function, whereas in 
Sec. \ref{secpf} we present our phase-field equations. We then show in 
Sec. \ref{seclim} that the phase-field equations reduce to the
macroscopic ones in the sharp-interface limit.
Deviations from that limiting behavior
are derived from the phase-field equations
themselves to 
first
order in the interface thickness in Sec. \ref{seccor}, and their
effect on the linear regime is computed in Sec. \ref{secrd}.
Finally, a brief summary is given in Sec. \ref{secdis}.
 
\section{The model}
\label{secmod}
We consider the general case of an interface with surface tension $\sigma$
between two fluids with distinct viscosities
($\mu_1$, $\mu_2$) and densities ($\rho_1$, $\rho_2$) moving in a
rectangular Hele--Shaw cell of width $W$ ($x$-direction) and gap $b$
($z$-direction),
under an effective
gravity $g_{eff}$ (negative $y$-direction)
and with an injection velocity $V_\infty$ (positive $y$-direction).
Label 1 (2) corresponds
to the upper (lower) fluid.

\subsection{Macroscopic equations}
\label{secmac}
Darcy's law is assumed to hold for each fluid, thus defining 
a certain velocity potential in each bulk,
but not on the interface. 
In contrast, the bulk incompressibility 
and the continuity of normal velocities on the interface
allow us to define its harmonic conjugate,
the stream function $\psi$, even on the interface through
$u_x=\partial_y \psi$, $ u_y=-\partial_x \psi$, where $u_x$, $u_y$ are the
$x,y$ components of the fluid velocity field $\vec u$.
Then Darcy's law results in a Laplace equation for the stream function
(potential flow) and a certain jump for the tangential fluid velocities
on the interface, whose value takes into account the 
Gibbs--Thomson relationship. The fact that the stream function is continuous 
at the interface makes the use of this variable particularly convenient. 
The Hele-Shaw equations in stream function formulation \cite{aref1} can be 
written in dimensionless 
form as:

\begin{equation}
\label{eq:laplace}
\nabla^2\psi=0 
\end{equation}
\begin{equation}
\label{eq:discontinuity}
\psi_r(0^+)-\psi_r(0^-)=-\gamma-c[\psi_r(0^+)+\psi_r(0^-)]
\end{equation}
\begin{equation}
\label{eq:continuity}
\psi_s(0^+)=\psi_s(0^-)=-v_n,
\end{equation}
which constitute the Hele--Shaw equations.
Here $r$ is a coordinate normal to the interface and
with origin on it, positive in fluid 1 ($0^\pm$ then
means on the interface coming from each side), $s$ is arclength along the
interface and such that the unit vectors satisfy $\hat s \times \hat r =
\hat x \times \hat y$, the subscripts 
stand for partial derivatives except for $v_n(s)$, which
is the normal velocity of the
interface, positive towards fluid 1, 
and 
\begin{equation}
\label{eq:gammasharp}
\frac{\gamma (s)}{2}\equiv B\kappa_s+\hat y \cdot \hat s,
\end{equation}
with
$\kappa (s)
$ the interface curvature, positive for a bump into fluid 2.
The 
dynamics are controlled by the two dimensionless parameters
\begin{equation}
\label{eq:bandc}
B=\frac{b^2\sigma}{12W^2[V_\infty(\mu_1-\mu_2)+g_{eff} \frac{b^2}{12}(\rho_1- \rho_2)]} 
\;\;\;,\;\;\;\;\; c=\frac{\mu_1-\mu_2}{\mu_1+\mu_2}.
\end{equation}
We will not be interested in negative values of $B$ (stable configuration) nor
$c$ (mirror image interface of $-c$).
So $B$ is a dimensionless surface tension,
and can be understood as the ratio between the capillary (stabilizing) force 
and the 
driving (destabilizing) force (injection+gravity), and $c$ is the viscosity
contrast, which is so far completely arbitrary: $0\le c\le 1$.
This corresponds to having set ourselves in the frame moving with the fluid
at infinity (or, equivalently, with the mean interface) and taken $W$ as unit
length
and $U_*\equiv cV_\infty+ g_{eff}
\frac{b^2(\rho_1-\rho_2)}{12(\mu_1+\mu_2)}$ as unit velocity
(see \cite{aref1}).


Note that Eqs. (\ref{eq:laplace},\ref{eq:discontinuity}) can be written
together as 
\begin{equation}
\label{eq:poisson}
\nabla^2\psi=-w, \;\;\;\; w=\{\gamma (s)+c[\psi_r(0^+)+\psi_r(0^-)]\}\delta(r)
\end{equation}
where $\delta(r)$ is the Dirac delta distribution and
$w\equiv \hat z\cdot (\vec\nabla
\times\vec u)$ is the fluid vorticity, which is confined to the interface.


\subsection{Phase-field equations}
\label{secpf}

We put forward the following phase-field model for the above equations
with $\theta$ being the phase field:

\begin{equation}
\label{eq:sf}
\epsilon \frac{\partial\psi}{\partial t}=\nabla^2\psi+c\vec \nabla \cdot 
(\theta \vec \nabla \psi)+\frac{1}{\epsilon} \frac{1}{2\sqrt 2}
\gamma(\theta )(1-\theta^2)
\end{equation}

\begin{equation}
\label{eq:pf}
\epsilon^2 \frac{\partial \theta}{\partial t}=f(\theta)+\epsilon^2\nabla^2\theta
+\epsilon^2 \kappa(\theta ) |\vec \nabla \theta |+\epsilon^2 \hat z \cdot 
(\vec \nabla \psi \times \vec \nabla \theta)
\end{equation}
where $f(\theta )\equiv \theta (1-\theta^2)$, and
$\frac{\gamma(\theta)}{2}\equiv \hat s(\theta)\cdot(B\vec\nabla
\kappa(\theta)
+\hat y)$, 
$\kappa(\theta)\equiv -\vec\nabla \cdot \hat r(\theta)$,
with 
$\hat r(\theta)\equiv \frac{\vec\nabla \theta}{|\vec\nabla \theta|}$
and $\hat s(\theta)\equiv \hat r(\theta) \times \hat
z$, 
together with the boundary condition

\begin{equation}
\label{eq:bc}
\theta(y\rightarrow \pm \infty)=\pm 1,
\end{equation}
so that $\theta=+1 (-1)$ corresponds to fluid 1 (2).
 $\gamma(\theta ) $,
$\kappa(\theta ) $ are functionals which 
generalize the magnitudes defined above for the interface, now to 
any level-set of the phase-field.

If we leave the two last terms aside, Eq. (\ref{eq:pf}) is the 
Cahn--Hilliard equation for a non-conserved order parameter or model A 
(without noise) in the 
classification of Ref. \cite{hh} of time-dependent
Ginzburg--Landau models. The field in this model is known to relax towards a
kink solution of a certain width in a short time scale, and then to evolve to
minimize the length of the effective interface according to Allen--Cahn law
(i.e. with normal velocity proportional to the local curvature).
The factor multiplying the laplacian has been choosen to be
$\epsilon^2$ for the kink
width to be ${\cal O}(\epsilon)$, so that $\epsilon$ can be considered the 
interface thickness, i.e., the small parameter in the asymptotic analysis that
will be performed in next section. 
On the other hand, the $\epsilon^2$ factor in the 
time derivative ensures that the relaxation towards the
kink is much faster than the evolution of the interface.
Notice that model A describes the relaxational 
dynamics of a non-conserved order parameter, whereas our problem is 
actually non-relaxational and strictly conserved (mass consevation and 
inmiscibility). The other two term in the phase-field equation will 
correct this apparent contradiction. 
In order to cancel out the local Allen-Cahn dynamics of the interface which 
is buit in model A, we add the term 
$\epsilon^2 \kappa(\theta ) |\vec \nabla \theta |$. It will be shown that 
such term cancels out Allen-Cahn law by giving rise, to leading order,
 to an identical 
contribution but with opposite sign.
%
With these elements so far, our phase-field relaxes to a kink profile 
located along an arbitrary interface which (if sufficiently smooth) 
remains almost completely
stationary, regardless of its shape.
This is because the dynamical effect of surface tension associated 
to the Ginzburg-Landau free energy 
has been removed (up to first order)
and the interface has not
yet been coupled to the fluid flow, represented by the stream function. This
coupling is achieved by adding the last term in Eq. (\ref{eq:pf}), which stands
for $-\epsilon^2 \vec u\cdot \vec \nabla \theta$ and thus sets the phase-field
---and therefore the interface--- in the frame moving with the fluid velocity
$\vec u$. This term restores the fully nonlocal dynamics of the Hele--Shaw 
model. In particular it yields the continuity of normal velocities
Eq. (\ref{eq:continuity}) and reintroduces surface tension, which is 
contained in the dynamical equation for the stream function through 
$\gamma(\theta ) $.

As for Eq. (\ref{eq:sf}), its right hand side is intended to reproduce 
Eq. (\ref{eq:poisson}), and therefore also
Eqs. (\ref{eq:laplace}) and (\ref{eq:discontinuity}). If the phase-field 
$\theta$ has a kink shape, $1-\theta^2$ is a peaked function which, when
divided by $\epsilon$, gives rise to the delta distribution for the vorticity.
However, this only accounts for the $\gamma$ in the weight of the delta.
The part proportional to the viscosity contrast $c$ must be put apart
as the $c\vec \nabla \cdot
(\theta \vec \nabla \psi)$ term because of the non-local character of
$\psi_r(0^+)+\psi_r(0^-)$. Finally,
 the time derivative is multiplied by
$\epsilon$ to recover the laplacian (and not diffusive) behavior of the
Hele--Shaw flow in the sharp-interface limit.

In spite of important differences, the proposed phase field equations
Eqs. (\ref{eq:sf},\ref{eq:pf}) contain certain similarities to the problem of a thermal plume
in a Hele--Shaw cell under gravity \cite{benamar}. In such a problem there
is only one fluid heated from the centre of the channel.
The heat diffuses towards the lateral walls, but the temperature profile is not
linear, since the fluid density and viscosity decrease with temperature, so that
the fluid in the middle of the channel raises because of buoyancy. 
As a result a so-called {plume} of hot fluid with a shape similar
to the Saffman--Taylor finger, with a stationary upwards velocity
and a width close to $1/2$ is formed. Outside the plume the fluid is colder, 
and the transition between the two zones is relatively abrupt,
so that one can think in terms
of an interface of a certain small thickness. Thus, the equation for the phase
field
Eq. (\ref{eq:pf}) could be thought as a diffusion equation for the temperature 
in a thermal plume. However, the available equations for that
problem hold only for the steady state \cite{benamar}, whereas our phase-field
model is intended to describe the whole dynamics. Generalization of the 
thermal plume equations to include the dynamics is not trivial for
non-vanishing viscosity contrast. As a matter of fact,
Ref. \cite{benamar} must restrict itself to low viscosity contrasts 
---as it is the case in thermal plumes---,
whereas we formulate the model for arbitrary viscosity contrast.
An interesting difference is the term
$\epsilon^2 \kappa(\theta ) |\vec \nabla \theta |$ cancelling out
Allen--Cahn law. The absence of that term in the thermal plume equations
does not
prevent the Hele--Shaw steady state equations to be recovered in the
sharp-interface limit because of the lower power of $\epsilon$ used in 
the $\vec u\cdot \vec \nabla \theta$ term, {but then} Allen-Cahn law arises
in the corrections at first order in the interface thickness. In contrast, by 
means of this $\epsilon^2 \kappa(\theta ) |\vec \nabla \theta |$ term we 
achieve
cancellation of the Allen-Cahn law even in such corrections, as we will see in
section
\ref{seccor}.
Finally, another major difference in the case of thermal plumes
is the absence of surface tension.
 




\section{Sharp-interface limit}
\label{seclim}

In order to analyze the small-$\epsilon$ behavior of the phase-field
equations, Eqs. (\ref{eq:sf},\ref{eq:pf}), 
we expand their fields in powers of $\epsilon$. 
The expected abrupt variations of these fields 
through the interface will make 
it necessary to perform two different expansions.
In the interface region (inner region) 
we rescale the differential operators appearing in these phase-field
equations by rewritting 
them in terms of the streched normal coordinate $\rho \equiv r/\epsilon $
(see Appendix). 
The expansions in the inner region will be matched order by order
in powers of $\epsilon$ to those in the outer region (in the bulk far from the
interface) where the coordinates are not rescaled. 
The outer and inner expansion are written respectively as 
\begin{equation}
\label{eq:outexp}
a(r,s,t)=a_0(r,s,t)+\epsilon a_1(r,s,t)+\epsilon ^2 a_2(r,s,t)+...
\end{equation}
\begin{equation}
\label{eq:inexp}
A(\rho,s,t)=A_0(\rho,s,t)+\epsilon A_1(\rho,s,t)+\epsilon ^2 A_2(\rho,s,t)+...
\end{equation}
where capital letters denote fields written in terms of the rescaled coordinate.
This results in the following
matching conditions:
\begin{eqnarray}
\label{eq:matching}
A_0(\rho ,s,t)&=&a_0(0^\pm,s,t) \nonumber \\
A_1(\rho ,s,t)&=&a_1(0^\pm,s,t)+\rho a_{0,r}(0^\pm,s,t) 
\;\;\;\;\;\;\;\;\;\;\;\;\;\;\;\;
\;\;\;\;\;\;\;\;\;\;\;\;\;\;\;\;\;\;\;\; {\rm as} \;\;\rho \rightarrow \pm \infty
\\
A_2(\rho ,s,t)&=&a_2(0^\pm,s,t)+\rho a_{1,r}(0^\pm,s,t)+
\frac {\rho ^2}{2}a_{0,rr}(0^\pm,s,t) \nonumber \\
&...& \nonumber 
\end{eqnarray}
And therefore:
\begin{eqnarray}
\label{eq:termmatch}
A_{0,\rho}(\pm\infty,s,t)&=&
A_{1,\rho\rho}(\pm\infty,s,t)=...=0, \nonumber \\
A_{1,\rho}(\pm\infty,s,t)&=&a_{0,r}(0^\pm,s,t) \\
A_{2,\rho}(\rho,s,t)&=&a_{1,r}(0^\pm,s,t)+\rho a_{0,rr}(0^\pm,s,t)
\;\; {\rm as} \;\;\rho \rightarrow \pm \infty \nonumber \\
&...& \nonumber
\end{eqnarray}

In practice, one does not find explicit solutions for the fields,
but some set of equations for them. 
A sharp-interface model for
the small-$\epsilon$ dynamics of the
phase-field
equations, Eqs. (\ref{eq:sf},\ref{eq:pf}), is then given by
the set of equations obeyed by the outer fields: Those obtained
at lowest order in the interface thickness $\epsilon$ (${\cal O}(\epsilon^0)$) 
constitute the $\epsilon \rightarrow 0$ limit 
of the phase-field model, which
we carry out in this section;
whereas
those obtained up to ${\cal O}(\epsilon)$ represent what we will
call (following Karma and Rappel
\cite{karma}) a `thin-interface' model, 
a model keeping finite interface thickness
effects, such as the one derived in Sec. \ref{seccor}.

\subsection{Outer equations}
Straightforward substitution of the outer expansion Eq. (\ref{eq:outexp})
in the outer equations, Eqs.
(\ref{eq:sf},\ref{eq:pf}), will yield the
bulk fields: a functional dependence for the phase-field and a differential
equation for the stream function.

Eq. (\ref{eq:pf}) at ${\cal O} (\epsilon^0)$ and ${\cal O} (\epsilon)$
reads respectively

\begin{equation}
\label{eq:pfout0}
{\cal O} (\epsilon_0): f_0(\theta)=f(\theta_0)=0 \Longrightarrow 
\theta_0=0,{\pm 1}=const 
\end{equation}
\begin{equation}
\label{eq:pfout1}
{{\cal O}} (\epsilon): f_1(\theta)=-2\theta_1=0 \Longrightarrow \theta_1=0,
\end{equation}
and iterating we get 
\begin{equation}
\label{eq:pfout}
\theta_i=0 \;\; \forall i>0
\end{equation}

Due to Eqs. (\ref{eq:pfout0}) and (\ref{eq:pfout}), $\theta=\pm 1$ to all orders,
and, 
therefore, the $(1-\theta^2)$ term in Eq. (\ref{eq:sf}) does not enter this
outer limit, whereas the viscosity contrast term in that equation
becomes $\pm c\nabla^2 \psi$,
depending on the phase. Hence, Eq. (\ref{eq:sf}) reads in this outer region
\begin{equation}
\label{eq:sfout}
\epsilon\frac{\partial \psi}{\partial t}=(1\pm c)\nabla^2\psi,
\end{equation}
which implies
\begin{equation}
\label{eq:sfout0}
\nabla^2\psi_0 = 0, \;\;\;\;\;\; 
\frac{\partial \psi_i}{\partial t}=(1\pm c)\nabla^2\psi_{i+1} \;\;\; \forall 
i\ge 0
\end{equation}
except for $c=1$. 
Note that we have recovered the sharp-interface Eq. (\ref{eq:laplace})
in the $\epsilon \rightarrow 0$ limit.
For $c=1$, Eq. (\ref{eq:laplace}) is still recovered in the +1 phase (
viscous fluid), whereas
in the -1 phase (inviscid fluid) the stream function turns out to be
constant in time 
to all orders. Although the inviscid fluid does not enter the problem in this
limit (see Eq. \ref{eq:discontinuity}), it still has a non-trivial dynamics,
since the stream function in it must evolve to keep satisfying
Eq. (\ref{eq:continuity}), and therefore, strictly speaking,
we do not really get the right
sharp-interface limit for $c$ exactly equal to one. However, 
the model can be 
applied to physical high viscosity contrast pairs of fluids. We shall come back
to this point in section \ref{seccor}.

\subsection{Inner equations}
\label{secin}
In turn, the interface boundary conditions for the stream function 
are given by the leading-order outer quantities $\psi_{0,s}(0^\pm)$ and 
$\psi_{0,r}(0^+)-\psi_{0,r}(0^-)$. According to the matching conditions
Eqs. (\ref{eq:matching}) and (\ref{eq:termmatch}), these equal the inner ones
$\Psi_{0,s}(\pm\infty)$ and
$\Psi_{1,\rho}(+\infty)-\Psi_{1,\rho}(-\infty)$ respectively. Because of the
specific structure of our phase-field equations, 
Eqs. (\ref{eq:sf},\ref{eq:pf}), we will need the first two orders in the inner
version of
Eq. (\ref{eq:pf}) and the lowest one in that of Eq. (\ref{eq:sf}) to get 
$\Psi_{0,s}(\pm\infty)$, and the two first in Eq. (\ref{eq:sf}) and the lowest
in Eq. (\ref{eq:pf}), to get $\Psi_{1,\rho}(+\infty)-\Psi_{1,\rho}(-\infty)$.
Therefore, we compute the two first orders in both Eqs. (\ref{eq:pf}) and 
(\ref{eq:sf}), by substituting the inner expansion Eq. (\ref{eq:inexp}) in the
inner (rescaled)
equations
(all whose terms are
derived in the appendix):

 Eq. (\ref{eq:pf})
up to ${\cal O} (\epsilon)$ reads
(see Eqs. (\ref{eq:inexp},\ref{eq:temps},\ref{eq:vectorial},\ref{eq:lapl+curv}))

\begin{equation}
\label{eq:pfin}
-\epsilon v_n \Theta_{0,\rho}=f(\Theta_0)+\epsilon \Theta_1 f'(\Theta_0)
+\Theta_{0,\rho \rho}+\epsilon \Theta_{1,\rho \rho}
+\epsilon (\Theta_{0,\rho} \Psi_{0,s}
-\Theta_{0,s} \Psi_{0,\rho}).
\end{equation}
Its ${\cal O} (\epsilon^0)$ part,
\begin{equation}
\label{eq:pfin0}
f(\Theta_0) +\Theta_{0,\rho \rho} = 0,
\end{equation}
which, together with the boundary conditions specified by the matching
(Eqs. (\ref{eq:matching},\ref{eq:termmatch})) with the outer expansion
 Eq. (\ref{eq:pfout0}), gives the so-called {kink} solution:
\begin{equation}
\label{eq:kink}
\Theta_0=\tanh \frac{\rho}{\sqrt 2}\;\;\;\;\Longrightarrow \;\;\;\;
\Theta_{0,\rho}=\frac{1}{\sqrt 2}{\rm sech} ^2 \frac {\rho }{\sqrt 2}=
\frac{1}{\sqrt 2}(1-\Theta_0^2)
\end{equation}
Hence we find the $\Theta_{0,s}$ term to vanish, and Eq. (\ref{eq:pfin}) reads at
${\cal O} (\epsilon)$:
\begin{equation}
\label{eq:pfin1}
  -v_n \Theta_{0,\rho}=\Theta_1 f'(\Theta_0)+\Theta_{1,\rho \rho}+\Theta_{0,\rho}
\Psi_{0,s}
\end{equation}

As for Eq. (\ref{eq:sf}), it reads, up to ${{\cal O}} (\frac{1}{\epsilon})$ 
(see Eqs. 
(\ref{eq:inexp},\ref{eq:laplaciana},\ref{eq:temps},\ref{eq:divbgrada},\ref{eq:gamma})),
\begin{eqnarray}
\label{eq:sfin}
& \frac {1}{\epsilon^2} & \Psi_{0,\rho \rho} +\frac {1}{\epsilon}
(\Psi_{1,\rho \rho}-\kappa \Psi_{0,\rho})
+c\{ \frac {1}{\epsilon^2}(\Theta_0 \Psi_{0,\rho})_\rho
   +\frac {1}{\epsilon}[(\Theta_0 \Psi_{1,\rho})_\rho
                        +(\Theta_1 \Psi_{0,\rho})_\rho
               -\kappa \Theta_0 \Psi_{0,\rho}]\} \nonumber \\
&+&\frac{1}{\epsilon}\frac{1}{2\sqrt 2}\gamma(1-\Theta_0^2)=0.
\end{eqnarray}
From its ${{\cal O}} (\frac {1}{\epsilon^2 })$ part we know that
\begin{equation}
\label{eq:sfin-2}
\Psi_{0,\rho}(1+c\Theta_0 )=const.
\end{equation}
Since $\Psi_{0,\rho}$ has no correspondence with the outer expansion, it must 
vanish at infinity (Eq. (\ref{eq:termmatch})). Then, we know the constant to be
zero. Now, since the term 
in brackets vanishes only for $c=1$, $\rho\rightarrow -\infty$, we deduce that 
\begin{equation}
\label{eq:psirho0}
\Psi_{0,\rho}=0  
\end{equation}
We then put Eq. (\ref{eq:psirho0}) into Eq. (\ref{eq:sfin}) at ${{\cal O}} (\frac
{1}{\epsilon })$:
\begin{equation}
\label{eq:sfin-1}
\Psi_{1,\rho \rho}+c(\Theta_0 \Psi_{1,\rho})_\rho=-\frac{\gamma}{2}\Theta_{0,\rho}
\end{equation}

Finally, Eqs. (\ref{eq:pfin1}) and (\ref{eq:sfin-1}) will yield the macroscopic
equations Eqs. 
(\ref{eq:continuity}) and (\ref{eq:discontinuity}) respectively:
Eq. (\ref{eq:pfin1}) can be rewritten in the form 

\begin{equation}
\label{eq:pfin1l}
\hat L \Theta_1\equiv[f'(\Theta_0)+\frac{\partial ^2}{\partial \rho^2}]\Theta_1=
-\Theta_{0,\rho
  } (v_n+\Psi_{0,s}).
  \end{equation}
We realize that $\hat L \Theta_{0,\rho}$ equals the partial
derivative with respect to $\rho$ of Eq. (\ref{eq:pfin0}), which, in turn,
vanishes.
Hence, we write down the solvability condition

\begin{equation}
\label{eq:intpfin1}
\int_{-\infty}^{+\infty}(v_n+\Psi_{0,s})\Theta_{0,\rho}^2  d\rho
=0
\end{equation}
Using Eq. (\ref{eq:psirho0}) we know that $\Psi_{0,\rho s}=0=\Psi_{0,s \rho }$ and
can take $\Psi_{0,s}$ 
out of the integral as well as $v_n$. Since the quantity left under the integral
sign $(\Theta_{0,\rho})^2$ 
is always positive, we find that $v_n+\Psi_{0,s}$ must vanish, and, matching with
the outer 
expansion, we get Eq. (\ref{eq:continuity}) for
$\psi_{0}$.

On the other hand, integrating Eq. (\ref{eq:sfin-1}) from $\rho \rightarrow
-\infty$
to $\rho \rightarrow +\infty$ we get
\begin{equation}
\label{eq:intsfin-1}
\Psi_{1,\rho}
=-\frac{\gamma}{2}\Theta_0-c\Theta_0\Psi_{1,\rho}+a_1(s),
\end{equation}
where $a_1(s)$ is an arbitrary function of $s$. Computing
$\Psi_{1,\rho}(+\infty)-\Psi_{1,\rho}(-\infty)$
and 
matching with the outer expansion
Eq. (\ref{eq:termmatch}) gives Eq. (\ref{eq:discontinuity}) for
$\psi_{0}$.
This completes the sharp-interface limit.

\section{First order corrections to the sharp-interface limit}
\label{seccor}


In the phase field model the interface width and the convergence to the sharp
interface limit is controlled by the small but finite value of the parameter
$\epsilon$.
Then, 
the question of which value of $\epsilon$ is needed to accurately reproduce
the actual Hele--Shaw dynamics for given values of the physical parameters $B$ 
and $c$ arises.
This question can be qualitatively answered by noting the distinct roles 
played by $\epsilon$ in the phase-field equations, Eqs. (\ref{eq:sf},\ref{eq:pf}):

The $\epsilon$ factors appearing in $\epsilon^2\nabla^2\theta$,
$\epsilon^2 \kappa(\theta ) |\vec \nabla \theta |$ and 
$\frac{1}{\epsilon} \frac{1}{2\sqrt 2}
\gamma(\theta )(1-\theta^2)$ all stand for the interface thickness, and this is
required to be small compared to the longitudinal length scale
$|k|^{-1}$ of the interface: $\epsilon |k|<<1$.

In contrast, the $\epsilon$ in 
$\epsilon \frac{\partial\psi}{\partial t}$ has nothing to do with the interface
thickness (and we will therefore denote it by $\tilde{\epsilon}$ from now on),
but its 
aim is to ensure that
the stream function is laplacian and not diffusive in the $\tilde{\epsilon}
\rightarrow 0$ limit, which commutes with the $\epsilon \rightarrow 0$ one
(the reader can convince himself of this by going through the limit again
but now considering $\tilde{\epsilon}$ of ${\cal O}(\epsilon^0)$):
$\tilde{\epsilon}$ sets the  
time scale of the diffusion of the stream function through a given
characteristic length of wavenumber $k$,
$\frac{\tilde{\epsilon}}{(1\pm c)k^2}$
(see Eq. \ref{eq:sfout}),
which must be much smaller than the characteristic growth rate of the interface
$|\omega|^{-1}$, so that the stream function is slaved to the interface:
 $\frac{\tilde{\epsilon}|\omega|}{k^2}<<1\pm c$.
We also realize that the viscosity contrast $c$
can be arbitrarily raised, as long as $\tilde{\epsilon}$ is correspondingly
lowered.
So our model is valid even for $c\rightarrow1$, as long as this limit
is taken formally after the $\epsilon \rightarrow 0$ one.

The $\epsilon^2$ in 
$\epsilon^2 \frac{\partial \theta}{\partial t}$ represents the
relaxation time of the phase field towards the steady kink solution (see Eq.
\ref{eq:pf}), 
which must be kept well 
below the interface growth time $|\omega|^{-1}$ for the phase-field to remain
close to the kink profile during the interface evolution:
$\epsilon^2|\omega|<<1$. This factor must be the same that the one in 
$\epsilon^2 \hat z \cdot
(\vec \nabla \psi \times \vec \nabla \theta)$ in order to get the macroscopic
equation Eq. (\ref{eq:continuity}). 
In fact  there are at least two distinct powers of $\epsilon$ for this relaxation
time ($\epsilon$ and $\epsilon^2$) for which the right sharp-interface limit is
achieved, and the corrections which we will 
compute would also be the same.

     To sum up, there are at least two independent small parameters
($\epsilon$ and $\tilde{\epsilon}$)
controlling the limit. When trying to approach macroscopic solutions
by means of numerical
integration of the phase-field equations,
it is very convenient to vary them independently in order to save computing
time, since both affect it \cite{numeric}.

A more quantitative answer to the question of the necessary values of
$\epsilon$, $\tilde{\epsilon}$ to get a given precision can be given by
extending the asymptotic analysis of the previous section to first order
in the interface thickness $\epsilon$ considering $\tilde{\epsilon}$ of
${\cal O}(\epsilon)$.
Thus, we will obtain a thin-interface model containing the corrections to the
limit up to that order in 
$\epsilon$ and $\tilde{\epsilon}$.

According to the matching conditions Eqs. (\ref{eq:matching}), 
the corrections to the interface boundary conditions for the stream function
at first order in $\epsilon$, $\psi_{1,s}(0^\pm)$ and
$\psi_{1,r}(0^+)-\psi_{1,r}(0^-)$, are to be identified as terms in the
expansion of $\Psi_{1,s}(\pm\infty)$ and
$\Psi_{2,\rho}(+\infty)-\Psi_{2,\rho}(-\infty)$ respectively. 
Now we will need the second order in Eq. (\ref{eq:pf}) and the first in
Eq. (\ref{eq:sf}) to compute $\Psi_{1,s}(\pm\infty)$, and the second in
Eq. (\ref{eq:sf}) and the first in Eq. (\ref{eq:pf}), to get 
$\Psi_{2,\rho}(+\infty)-\Psi_{2,\rho}(-\infty)$. Therefore, we must compute the 
next order both in Eqs. (\ref{eq:pf}) and (\ref{eq:sf}), but, first, we can still 
extract some information from the lower orders.

On the one hand, we found that $\Psi_{0,s}=-v_n$. We put this into
Eq. (\ref{eq:pfin1}) to get the
differential equation for $\Theta_1$:
\begin{equation}
\label{eq:pfin11}
\Theta_1 f'(\Theta_0)+\Theta_{1,\rho \rho}=0
\end{equation}
with boundary conditions coming from the matching Eq. (\ref{eq:matching}) with
Eq. (\ref{eq:pfout})
$\Theta_1(\pm \infty)=\Theta_{1,\rho}(\pm \infty)=0$
and solution $\Theta_1=0$.

The integral with respect to $\rho$ of Eq. (\ref{eq:intsfin-1})
is
\begin{equation}
\label{eq:int2sfin-1}
\Psi_1=-\frac{\gamma}{2}\int\frac{\Theta_0 d\rho}{1+c\Theta_0}+a_1(s)
\int\frac{d\rho}{1+c\Theta_0}.
\end{equation}
According to the matching Eq. (\ref{eq:matching}), 
the $\rho \rightarrow \pm \infty$ asymptotics of $\Psi_1(\rho)$
should consist of a finite term, $\psi_1(0^\pm)$, and a diverging one,
$\rho \psi_{0,r}(0^\pm)$.
For vanishing viscosity contrast the last integral in Eq. (\ref{eq:int2sfin-1})
is $\rho a_1(s)$ and
clearly does not contribute to the finite term $\psi_1(0^\pm)$. Then, since
$\Theta_0$ is an odd function of $\rho$, its integral with respect
to $\rho$ will
be even, and $\psi_1(0^+)$=$\psi_1(0^-)$, i.e., the fluid velocity normal to
the interface will be continous on it. For non-zero values of c, however, one
must compute the integrals in Eq. (\ref{eq:int2sfin-1}), find their 
$\rho \rightarrow \pm \infty$ asymptotics, and identify $\psi_1(0^\pm)$
and $\psi_{0,r}(0^\pm)$. Requiring this latter quantity to be consistent with
Eq. (\ref{eq:discontinuity}) for $\psi_0$, one fixes $a_1(s)$, and putting this
back into the
identified $\psi_1(0^\pm)$ value, one finds
\begin{equation}
\label{eq:psi1}
\psi_1(0^\pm)=-\frac{\sqrt 2}{2}\{\gamma+c[\psi_{0,r} (0^+)+\psi_{0,r}(0^-)]\}
\ln
\frac{1\pm c}{2} \;+a_2(s) 
\end{equation}
where $a_2(s)$ is another arbitrary function of $s$. This will give rise to a
discontinuity in the fluid velocity:
\begin{eqnarray}
\label{eq:discvel}
\psi_{1,s}(0^+)-\psi_{1,s}(0^-)&=&-\frac{\sqrt 2}{2}\{\gamma_s
+c[\psi_{0,rs} (0^+)+\psi_{0,rs}(0^-)]\}\ln\frac{1+c}{1-c}= \nonumber \\
&=&
-c\sqrt 2 \{\gamma_s
+c[\psi_{0,rs} (0^+)+\psi_{0,rs}(0^-)]\}+{\cal O}(c^3)
\end{eqnarray}

In order to fix $\partial_s a_2(s)$, we compute the next order $({\cal
O}(\epsilon^2))$ of Eq. (\ref{eq:pf}) to get
(see Eqs. 
(\ref{eq:inexp},\ref{eq:temps},\ref{eq:vectorial},\ref{eq:lapl+curv})):
\begin{equation}
\label{eq:pfin2}
\Theta_2f'(\Theta_0)+\Theta_{2,\rho \rho}
-\rho\kappa v_n \Theta_{0,\rho}+\Psi_{1,s}\Theta_{0,\rho}=0
\end{equation}
This has the same structure than Eq. (\ref{eq:pfin1}) and an analogue solvability
condition applies:

\begin{equation}
\label{eq:intpfin2}
\int_{-\infty}^{+\infty} \Psi_{1,s}\Theta_{0,\rho}^2d\rho=0
\end{equation}
Substitution of the expression for $\Psi_1$ obtained by performing the
integrals in
Eq. (\ref{eq:int2sfin-1}) into this condition and subsequent computation of the
resulting integral fixes $\partial_s a_2(s)$ so that

\begin{eqnarray}
\label{eq:psi1s}
\psi_{1,s}(0^\pm)&=&
-\frac{\sqrt 2}{2}[\frac{\gamma_s}{2}
+c\frac{\psi_{0,rs} (0^+)+\psi_{0,rs}(0^-)}{2}]
[1-\frac{1}{c^2}+(\pm 1+\frac{\frac{1}{c^3}-\frac{3}{c}}{2})\ln\frac{1+c}{1-c}]=
\nonumber \\
&=&{\sqrt 2}[\frac{\gamma_s}{2}
+c\frac{\psi_{0,rs} (0^+)+\psi_{0,rs}(0^-)}{2}]
[\frac{5}{6}\mp c+\frac{2}{5}c^2+{\cal O}(c^3)]
\end{eqnarray}

Finally, to get $\Psi_{2,\rho}(+\infty)-\Psi_{2,\rho}(-\infty)$ we need
Eq. (\ref{eq:sf}) at $({\cal O}(\epsilon^0))$ 
(see Eqs. 
(\ref{eq:inexp},\ref{eq:laplaciana},\ref{eq:temps},\ref{eq:divbgrada},\ref{eq:gamma})):
\begin{equation}
\label{eq:sfin0}
\Psi_{2,\rho \rho}-\kappa \Psi_{1,\rho}-\partial_s v_n
+c[(\Theta_0\Psi_{2,\rho})_\rho-\kappa\Psi_{1,\rho}\Theta_0-\partial_s v_n
   \Theta_0]
+3B\rho\kappa\kappa_s\Theta_{0,\rho}=0
\end{equation}
Integrating this from $\rho \rightarrow -\infty$ to
$\rho \rightarrow +\infty$ we obtain
\begin{equation}
\label{eq:intsfin0}
[\Psi_{2,\rho }]_{-\infty}^{+\infty}
-\kappa\int_{-\infty}^{+\infty}(1+c\Theta_0)\Psi_{1,\rho}d\rho
-\partial_s v_n[\rho]_{-\infty}^{+\infty}
+c[\Theta_0\Psi_{2,\rho }]_{-\infty}^{+\infty}=0
\end{equation}
where we have omitted integrals of odd functions of $\rho$. We use
Eq. (\ref{eq:intsfin-1}) to rewrite the integrand of the remaining integral
as $-\frac{\gamma}{2} \Theta_0+a_1(s)$. $\Theta_0$ is an odd function of $\rho$
and
does not contribute to the integral, whereas $a_1(s)$ gives rise to a divergent
term of the type $[\rho]_{-\infty}^{+\infty}$. According to the matching
Eq. (\ref{eq:termmatch}), $\psi_{1,r}(0^\pm)$ corresponds to the finite part of 
$\Psi_{2,\rho }(\pm \infty)$, so that we find
\begin{equation}
\label{eq:psi1r}
\psi_{1,r}(0^+)-\psi_{1,r}(0^-)=-c[\psi_{1,r}(0^+)+\psi_{1,r}(0^-)]
\end{equation}
which will leave the jump of the normal derivative of the stream function 
across the interface unaffected at first order in the kink width.

Putting Eqs. 
(\ref{eq:sfout},\ref{eq:continuity} and \ref{eq:discontinuity} for
$\psi_0$,\ref{eq:psi1s},\ref{eq:psi1r})
together, we get an effective sharp-interface
model for the dynamics of the $\theta=0$ level-set up to first order in 
$\epsilon$ and $\tilde{\epsilon}$:
\begin{eqnarray}
\label{eq:difusio}
\tilde{\epsilon}\frac{\partial \psi}{\partial t}&=&(1\pm c)\nabla^2\psi
\\
\label{eq:psir}
\psi_r(0^+)-\psi_r(0^-)&=&-\Gamma \\
\label{eq:psis}
\psi_s(0^\pm)&=&
-v_n
-\epsilon\frac{\sqrt 2}{2}\frac{\Gamma_s}{2}g_\pm(c)
\;\;\;\;\nonumber \\
&=&-v_n+\epsilon{\sqrt 2}\frac{\Gamma_s}{2}
[\frac{5}{6}\mp c+\frac{2}{5}c^2+{\cal O}(c^3)], 
\end{eqnarray}
where $\Gamma\equiv\gamma+c[\psi_r(0^+)+\psi_r(0^-)]$ is the weight of the 
vorticity defined in Eq. (\ref{eq:poisson}) evaluated up to ${\cal O}(\epsilon)$
and $g_\pm(c)=
1-\frac{1}{c^2}+(\pm 1+\frac{\frac{1}{c^3}-\frac{3}{c}}{2})\ln\frac{1+c}{1-c}$.

Note that the desired corrections to the limiting equations 
Eqs. (\ref{eq:laplace}-\ref{eq:continuity}) in Eqs. (\ref{eq:difusio}) and
(\ref{eq:psis}) go as 
$\tilde{\epsilon}$ and $\epsilon$ respectively, and the fact that 
Eq. (\ref{eq:discontinuity}) remains unaffected. Note as well that the correction
in $\epsilon$ appearing in Eq. (\ref{eq:psis})
has nothing to do with an Allen--Cahn law. So the
$\epsilon^2 \kappa(\theta ) |\vec \nabla \theta |$ term 
has cancelled this out even in the first order corrections. 

\section{Linear dispersion relation up to
first order in the interface thickness}
\label{secrd}

In order to see how such corrections affect some relevant specific situation
we compute the linear dispersion relation of a perturbation to the planar 
interface $y(x)=Ae^{\omega t+ikx}$ for Eqs. (\ref{eq:difusio}-\ref{eq:psis}).
We make the ansatz

\begin{equation}
\label{eq:psilineal}
\psi(x,y)=a_\pm Ae^{\omega t+ikx-q_\pm |y|}
\end{equation}
inspired by the actual Hele--Shaw result, where now the coefficient
$a_\pm$ allows for distinct amplitudes in each phase for the stream function
to satisfy the discontinuity in the normal velocities of Eq. (\ref{eq:psis}),
whereas the decay length $q_\pm$ in the y-direction is set not 
only by the wavelength of the perturbation $2\pi/k$, but also by the diffusion
length in Eq. (\ref{eq:difusio}), which is also different in each phase.
Thus, Eq. (\ref{eq:difusio}) yields
\begin{equation}
\label{eq:q}
q_\pm=|k|p_\pm, \;\;\;\;
p_\pm=+\sqrt{1+\frac{\tilde{\epsilon}\omega}{k^2(1\pm c)}}
\end{equation}
In turn, taking into account that $v_n=\omega Ae^{\omega t+ikx}$ and
$\gamma=2iA{\rm sign}(k)\omega_0e^{\omega t+ikx}$
---where $\omega_0=|k|(1-Bk^2)$ is the actual Hele--Shaw growth rate---,
Eq. (\ref{eq:psis}) fixes $a\pm$ to be

\begin{equation}
\label{eq:a}
a_\pm=\frac{i\omega}{k}[1+\epsilon|k|{\sqrt 2}g_\pm (c)\frac{p_+ +p_-}{2}]
\end{equation}
Finally, Eq. (\ref{eq:psir}) requires that the following dispersion relation is
satisfied:

\begin{eqnarray}
\label{eq:w}
\omega&=&\frac{\omega_0}
{\frac{(1+c)p_- +(1-c)p_+}{2}}
[1-\epsilon|k|{\sqrt 2}\frac{p_- +p_+}{2}\frac{g_+ (1-c)p_+ +g_- (1+c)p_- }
{(1-c)p_+ +(1+c)p_- }]+{\cal O}(\epsilon^2)  \\
\label{eq:w1}
&=&\omega_0(\frac{1}{\sqrt{1+\frac{\tilde{\epsilon}\omega}{k^2}}}
-\epsilon|k|{\sqrt 2}\frac{5}{6})+{\cal O}(c^2)+{\cal O}(\epsilon^2)
\end{eqnarray}
This consists of the well known Hele--Shaw growth rate multiplied
by a factor smaller than 1 carrying the corrections in $\epsilon$,
$\tilde{\epsilon}$. We identify
the conditions on
$\epsilon$,
$\tilde{\epsilon}$ heuristically derived at the beginning of
section \ref{seccor} to control
how close this factor is to 1 and in general how close the stream function is to
the actual Hele--Shaw one: $\tilde{\epsilon}\omega/k^2<<1\pm c$ (within $p_\pm$)
and $\epsilon |k|<<1$ in Eqs. (\ref{eq:q}-\ref{eq:w}); and the
simplified version
up to ${\cal O}(c)$ $\tilde{\epsilon}\omega/k^2<<1$ and $\epsilon |k|<<1$
in Eq. (\ref{eq:w1}). The amplitude factor Eq. (\ref{eq:a}) can also be expanded
in
powers of $c$ making use of Eq. (\ref{eq:w1}) to find
\begin{equation}
\label{eq:a1}
a_\pm =\frac{i\omega_0}{k}(\frac{1}{\sqrt{1+\frac{\tilde{\epsilon}\omega}{k^2}}}
\mp c\epsilon|k|{\sqrt 2})+{\cal O}(c^2)+{\cal O}(\epsilon^2).
\end{equation}

Since these corrections have a stabilizing effect, they could affect the
selection of the steady finger width. As a matter of fact, Ben Amar already
showed that the $\partial_s (\hat y \cdot \hat s)$ term of $\Gamma_s$ in Eq.
(\ref{eq:psis})
on its own was capable of selecting a finger width greater than $1/2$
\cite{benamar}. Then, for small enough values of the physical surface tension
(i.e., the physical selection mechanism),
for which a width very close to $1/2$ should be expected, this term could turn
out to control the selection itself, so that an unexpected greater width
could be obtained.  
Of course, this will not be the case if a sufficiently small value of the 
interface thickness $\epsilon$ is used, so that the condition $\epsilon|k|<<1$ 
is satisfied for the length scale set by the surface tension:
$\epsilon<<\sqrt B$.

\section{Conclusions}
\label{secdis}
We have introduced a phase-field model for Hele--Shaw flows 
with arbitrary viscosity contrast and shown it to yield the proper
sharp-interface limit.
%
We have actually found two independent small parameters
($\epsilon$ and $\tilde{\epsilon}$) and three
distinct conditions on them to control the convergence to the 
sharp-interface limit $\epsilon,\tilde{\epsilon} \rightarrow 0$.
In particular, $\tilde{\epsilon}$ must be lowered when $c$ is
increased.
A thin-interface model, i.e. an effective sharp interface model keeping 
finite-$\epsilon$ and -$\tilde{\epsilon}$ effects,
has been derived for the dynamics of the phase-field model
up to first order in both of these parameters. This thin-interface
model has then been used to explicitely compute the 
finite-$\epsilon$ and -$\tilde{\epsilon}$ corrections to the
Hele--Shaw result for a specific situation such as the linear 
regime, thus suggesting that the single-finger width selection
could also be affected by these finite-thickness effects.

In the following paper \cite{numeric} we perform numerical simulations of the
phase field model Eqs. (\ref{eq:sf},\ref{eq:pf}), and we explicitly vary the two
small parameters $\epsilon$ and $\tilde{\epsilon}$ independently.
In this way we both
control the simulation
accuracy through the conditions mentioned to show how to reproduce
the Hele--Shaw dynamics within this method, and explicitly check
convergence in the interface thickness.

\section*{Acknowledgements}
We are indebted to J-L. Mozos, E. Corvera and
H. Guo
for collaboration in the early stages of this work.
We acknowledge financial support from the Direcci\'on
General de Ense\~{n}anza Superior (Spain), Projects No. PB96-1001-C02-02,
PB96-0378-C02-01 and PB96-0241-C02-02,
and the European Commission Project No. ERB FMRX-CT96-0085.
 R.F. also aknowledges
a grant from the Comissionat per a Universitats i Recerca
(Generalitat de Catalunya).

\appendix
\section*{}
Our goal here is to rescale the differential operators appearing in the phase
field model Eqs.
(\ref{eq:sf},\ref{eq:pf}).
The first step will be to rewrite them in terms of the local coordinates defined
on the interface $r$ and $s$.
To do this, one must precisely define the curvilinear coordinate
system and compute its so-called scale factors:

Consider the $\theta=0$ level-set and its intrinsic coordinates
$s$ (arclength along it) and $r$ (signed distance to it, positive for a point
with $\theta>0$), so that $\hat s \times \hat r=\hat x \times \hat y$. Let 
$\alpha $ be the angle going from $\hat x$ to $\hat s$. Then 
$\kappa= \alpha_s$ is the $\theta=0$ level-set curvature.
We introduce 
$X,Y$ as the values of $x,y$ for a point on the $\theta=0$ curve with
a given value of $s$. By moving this point
infinitessimaly along $s$ we find that these values have changed in
$dY=ds\sin\alpha$, $dX=ds\cos\alpha$. Consider also the coordinates of a
point $x,y$ with $\theta \neq 0$ in terms of the values $X,Y$ of its closest
neighbour on the $\theta=0$ level-set and the signed distance between them.
Taking into account that $\alpha$ is also the angle going from $\hat y$
to $\hat r$, one finds $x=X-r\sin \alpha$, $y=Y+r\cos\alpha$.
Now one can compute the (positive defined) scale factors

\begin{eqnarray}
\label{eq:hr}
h_r^2&\equiv&x_r^2+y_r^2=1 \Longrightarrow h_r
 =1 \\
 \label{eq:hs}
h_s^2&\equiv&x_s^2+y_s^2=(X_s-r\alpha_s\cos\alpha)^2+(Y_s-r\alpha_s\sin\alpha)^2= 
\nonumber \\
&=&(\cos\alpha-r\kappa\cos\alpha)^2+(\sin\alpha-r\kappa\sin\alpha)^2=
(1-r\kappa )^2 \Longrightarrow h_s
=|1-r\kappa|=1-r\kappa  
\end{eqnarray}
Note that the last equality in Eq. (\ref{eq:hs}) requires that $r\kappa<1$.
In the inner region, where we make use of such formulae, 
this will hold as long as the interface thickness $\epsilon$ is much smaller than
the curvature radius at any point of the interface, i.e. not too far from 
the sharp-interface limit. Otherwise the present analysis would
break down, because one could always find a point such that $r\kappa=1$, where
$h_s$ would vanish, reflecting the fact that the change of coordinates has
become
ambigous in $s$. 

Then, the scale factors are used to express the differential operators in terms
of $r$, $s$:

\begin{eqnarray}
\label{eq:gradout}
\vec\nabla a&=&a_r \hat r+\frac{a_s}{1-r\kappa} \hat s \\
\label{eq:divout}
\vec\nabla \cdot \vec a&=&(a^r)_r+\frac{-\kappa a^r+(a^s)_s}{1-r\kappa} 
\;\;\;\;\;\;\;\;\;\;\;\;\;\;\;\;
(\vec a =a^r \hat r+a^s \hat s) \\
\label{eq:laplacianaout}
\nabla^2 a&=&a_{rr}-\frac{\kappa a_r}{1-r\kappa}+\frac{a_{ss}}{(1-r\kappa)^2}
+\frac{r\kappa_s a_s}{(1-r\kappa)^3}
\end{eqnarray}

Finally, one sets $r=\epsilon\rho$ and
expands in powers of $\epsilon$:

\begin{eqnarray}
(1-r\kappa)^{-1}=1+\epsilon\rho\kappa+{\cal O}(\epsilon^2)
\end{eqnarray}
One gets:

\begin{eqnarray}
\label{eq:grad}
\vec\nabla a&=&\frac{1}{\epsilon}A_\rho \hat r +\hat s A_s
[1+\epsilon \rho \kappa+{\cal O}(\epsilon^2)] \\
\label{eq:div}
\vec\nabla \cdot \vec a&=&\frac{1}{\epsilon}(A^r)\rho
+[(A^s)_s-\kappa A^r][1+\epsilon \rho \kappa+{\cal O}(\epsilon^2)] \\
\label{eq:laplaciana}
\nabla^2 a&=&\frac{1}{\epsilon^2}A_{\rho\rho}-\frac{1}{\epsilon}\kappa A_\rho
-\rho\kappa^2A_\rho+A_{ss}+{\cal O}(\epsilon)
\end{eqnarray}
This completes the rescaling. Capital letters denote fields written 
in the rescaled coordinates of the inner region.
Any other quantity appearing in Eqs. (\ref{eq:sf},\ref{eq:pf}) is derived from
these ones. For instance we get:

\begin{equation}
\label{eq:temps}
\frac{\partial a}{\partial t}
\equiv \frac{\partial a}{\partial t}|_{x,y={\rm const}}
=\frac{da}{dt}-\vec v \cdot \vec\nabla a=
-\frac{v_n}{\epsilon}A_\rho+
\frac{da}{dt}
-v_t A_s+{\cal O}(\epsilon) 
\end{equation}
where the partial (total) time derivative is computed keeping $x,y$ 
($r,s$) fixed, and
$\vec v$ is the velocity of the $r,s$ frame respect to the $x,y$ one, i.e., the
interface velocity.
Moreover
\begin{eqnarray}
\label{eq:vectorial}
\hat z \cdot (\vec\nabla \psi \times \vec\nabla \theta)&=&
\frac{1}{\epsilon}\{\Psi_s\Theta_\rho [1+\epsilon \rho \kappa
+{\cal O}(\epsilon^2)]-\Psi_\rho \Theta_s[1+{\cal O}(\epsilon)]\} \\
\label{eq:divbgrada}
\vec\nabla \cdot(\theta\vec\nabla \psi)&=&\frac{1}{\epsilon^2}(\Theta
\Psi_\rho)_\rho
+(\Theta \Psi_s)_s
-\frac{1}{\epsilon}\kappa \Theta \Psi_\rho(1+\epsilon\rho\kappa)
+{\cal O}(\epsilon)= \nonumber \\
&=& \frac {1 } { \epsilon ^2 } (\Theta \Psi_\rho)_\rho
-\frac{1}{\epsilon}\kappa \Theta \Psi_\rho-\rho\kappa^2\Theta \Psi_\rho
+(\Theta \Psi_s)_s +{\cal O}(\epsilon)
\Psi_\rho]+{\cal O}(\epsilon^2)
\end{eqnarray}

The only terms left in
Eqs. (\ref{eq:sf},\ref{eq:pf})
to compute are those containing $\gamma(\theta)$ and 
$\kappa(\theta)$. To construct them we will need the following quantities:
\begin{eqnarray}
\label{eq:mgrad}
|\vec\nabla \theta |&=&+\sqrt{\frac{1}{\epsilon^2}\Theta _\rho^2+\Theta _s^2
[1+{\cal O}(\epsilon)]^2}=
+\frac{\Theta _\rho}{\epsilon}\sqrt{1+\epsilon^2\frac{\Theta _s^2}
{\Theta _\rho^2}[1+{\cal O}(\epsilon)]^2}=
\nonumber \\
&=&+\frac{\Theta _\rho}{\epsilon}[1+\frac{\epsilon^2}{2}\frac{\Theta _s^2}
{\Theta _\rho^2}+{\cal O}(\epsilon^3)]
=\frac{\Theta _\rho}{\epsilon}+\frac{\epsilon}{2}\frac{\Theta _s^2}
{\Theta _\rho}+{\cal O}(\epsilon^2)
\end{eqnarray}
(Note that $\Theta_\rho>0$, since $\Theta$ is monotonic in $\rho$ 
and we defined $r$ to be positive for the $\theta>0$ phase)
\begin{eqnarray}
\label{eq:r}
\hat r(\theta)\equiv \frac{\vec\nabla \theta}{|\vec\nabla \theta|} & = & 
\frac{\frac{\epsilon}{\Theta _\rho}}{\frac{\epsilon}{\Theta _\rho}} \times 
\frac{\vec\nabla \theta}{|\vec\nabla \theta|}=
\frac{\hat r+\hat s\epsilon\frac{\Theta _s}{\Theta _\rho}[1+{\cal O}(\epsilon)]}
{1+\epsilon^2\frac{1}{2}\frac{\Theta _s^2}{\Theta _\rho^2}+{\cal O}(\epsilon^3)}
%
\nonumber \\
&=&\hat r[1-\frac{\epsilon^2}{2}\frac{\Theta _s^2}{\Theta _\rho^2}
+{\cal O}(\epsilon^3)]
+\hat s[ \epsilon\frac{\Theta _s}{\Theta _\rho}+{\cal O}(\epsilon^2)]
\end{eqnarray}
We have termed this $\hat r(\theta)$ because it is indeed the unit vector normal
to the
$\theta=$const level-set on which it is computed.
We similarly define $\hat s(\theta)\equiv \hat r(\theta) \times \hat z$
and 
$\kappa(\theta)\equiv -\vec\nabla \cdot \frac{\vec\nabla \theta}{|\vec\nabla
\theta|}$
generalized
to construct
$\frac{\gamma(\theta)}{2}\equiv B\hat s(\theta)\cdot\vec\nabla\kappa(\theta)
+\hat y \cdot \hat s(\theta)$:

\begin{eqnarray}
\label{eq:curv}
-\kappa(\theta)\equiv \vec\nabla \cdot \frac{\vec\nabla \theta}
{|\vec\nabla \theta|}&=& 
- \frac{\epsilon}{2}(\frac{\Theta _s^2}{\Theta _\rho^2})_\rho
+\epsilon(\frac{\Theta _s}{\Theta _\rho})_s
-\kappa(1+\epsilon \rho \kappa)+{\cal O}(\epsilon^2)
= \nonumber \\
&=&-\kappa+\epsilon[(\frac{\Theta _s}{\Theta _\rho})_s
-\frac{1}{2}(\frac{\Theta _s^2}{\Theta _\rho^2})_\rho
-\rho \kappa^2]+{\cal O}(\epsilon^2) \\
\label{eq:gradcurv}
\vec\nabla\kappa(\theta)=&-&\hat r[(\frac{\Theta _s}{\Theta _\rho})_{s\rho}
-\frac{1}{2}(\frac{\Theta _s^2}{\Theta _\rho^2})_{\rho\rho}
-\kappa^2+{\cal O}(\epsilon)]  \nonumber \\
&+&\hat s\{\kappa_s+\epsilon[-(\frac{\Theta _s}{\Theta _\rho})_{ss}
+\frac{1}{2}(\frac{\Theta _s^2}{\Theta _\rho^2})_{\rho s}
+2\rho \kappa\kappa_s]+{\cal O}(\epsilon^2)\}
[1+\epsilon\rho\kappa+{\cal O}(\epsilon^2)]=
 \nonumber \\
=&-&\hat r[-\kappa^2+(\frac{\Theta _s}{\Theta _\rho})_{s\rho}
-\frac{1}{2}(\frac{\Theta _s^2}{\Theta _\rho^2})_{\rho\rho}
+{\cal O}(\epsilon)]  \nonumber \\
&+&\hat s\{\kappa_s+\epsilon[3\rho \kappa\kappa_s
-(\frac{\Theta _s}{\Theta _\rho})_{ss}
+\frac{1}{2}(\frac{\Theta _s^2}{\Theta _\rho^2})_{\rho s}]
+{\cal O}(\epsilon^2)\} \\
\label{eq:gamma}
\frac{\gamma(\theta)}{2}
=B\{&\kappa_s&+\epsilon[3\rho \kappa\kappa_s
-(\frac{\Theta _s}{\Theta _\rho})_{ss}
+\frac{1}{2}(\frac{\Theta _s^2}{\Theta _\rho^2})_{\rho s}] \nonumber \\
&+&\epsilon\frac{\Theta _s}{\Theta _\rho}
[-\kappa^2+(\frac{\Theta _s}{\Theta _\rho})_{s\rho}
-\frac{1}{2}(\frac{\Theta _s^2}{\Theta _\rho^2})_{\rho\rho}]\} 
\nonumber \\
+\hat y &\cdot& \hat s -\epsilon\frac{\Theta _s}{\Theta _\rho}\hat y \cdot \hat
r
+{\cal O}(\epsilon^2)= \frac{\gamma}{2}+{\cal O}(\epsilon)
\end{eqnarray}
We should still compute the product 
$\kappa(\theta)|\vec\nabla\theta|$ appearing in Eq. (\ref{eq:pf}), but
instead we prefer
to compute straightahead the sum
$\nabla^2\theta+\kappa(\theta)|\vec\nabla\theta|=\nabla^2\theta
-\nabla^2\theta+\frac{ \vec\nabla\theta}{|\vec\nabla\theta|}
\cdot\vec\nabla|\vec\nabla\theta|=\hat r(\theta)
\cdot\vec\nabla|\vec\nabla\theta|$:

\begin{eqnarray}
\label{eq:gradmod}
\vec\nabla|\vec\nabla\theta|&=&
\hat r[\frac{\Theta_{\rho\rho}}{\epsilon^2}+\frac{1}{2}
(\frac{\Theta_s^2}{\Theta_\rho})_\rho+{\cal O}(\epsilon)] 
+\hat s[\frac{\Theta_{\rho s}}{\epsilon}+{\cal O}(\epsilon^0)] \\
\label{eq:lapl+curv}
\nabla^2\theta+\kappa(\theta)|\vec\nabla\theta|&=&
\frac{\Theta_{\rho\rho}}{\epsilon^2}
+\frac{1}{2}[(\frac{\Theta_s^2}{\Theta_\rho})_\rho
-\frac{\Theta_s^2}{\Theta_\rho^2}\Theta_{\rho\rho}
+2\frac{\Theta_s}{\Theta_\rho}\Theta_{\rho s}]+{\cal O}(\epsilon)
\nonumber \\
&=&\frac{\Theta_{\rho\rho}}{\epsilon^2}
+\frac{\Theta_s\Theta_{\rho s}}{\Theta_\rho}(1+\frac{1}{\Theta_\rho})
-\frac{\Theta_s^2\Theta_{\rho \rho}}{\Theta_\rho^2}
+{\cal O}(\epsilon)
\end{eqnarray}

\end{document}